\documentclass[twocolumn,twoside,preprintnumbers,amsmath,amssymb,showkeys]{revtex4}
\usepackage{epsfig}
\usepackage{graphicx}

\usepackage{fancyhdr}

\usepackage{pslatex}

\begin{document}

\title{Production of light particles by very strong and slowly varying magnetic fields}

\author{Giorgio Calucci$^1$, and Antonino Di Piazza$^2$}

\affiliation{$^1$ Dipartimento di Fisica teorica, Universit\`a di Trieste ed
INFN Sezione di Trieste, Italy \\
         $^2$ Max Planck Institut f\"ur Kernphysik 
     Heidelberg - Germany  }


\begin{abstract}

The possibility that around some astrophysical objects there are non-static
magnetic fields of enormous intensity suggests that in these situations real
particles may be produced. The slowness of the variation is compensated by the
huge intensity. The main issue is the production of $e^+-e^-$ pairs annihilating
into photons and the direct production of photons, as one of the concurrent
process in the GRB (gamma ray bursts). Then some simple effects due to the presence of the intense gravity are studied and finally a look is given to the
production of other kinds of particles.
\end{abstract}

\maketitle

\thispagestyle{fancy}

\setcounter{page}{1}

\section{MOTIVATIONS }

  There are phenomenological and formal motivations.\par
 For the
  first instance: there are the {\it Gamma-ray Bursts} and the
  signal is obviously electromagnetic: this electromagnetic signal could well be the final outcome of dynamical processes where other kinds of interaction are involved, but  we can look also for a direct electromagnetic origin of the phenomenon.
 We don't say that what we propose is the main
 mechanism, but this kind of mechanism must exist
 since it is a direct consequence of standard
 electrodynamics, provided we accept the existence of huge,
 slowly varying magnetic fields.\par 
 The formal motivation is that this kind of analysis yields an example of nonperturbative QED: It happens that, contrary to the very numerous and extremely accurate calculations in perturbative QED, the sector of nonperturbative QED is less frequently explored.

          \section{{GENERAL FEATURES}}

 We have to deal with a typical two-scale problem: there is the astrophysical scale and the elementary-particle scale. The astrophysical scale is characterized by low frequencies, but these low frequencies can be compensated by huge intensities of the magnetic field. We make more quantitative this statement:
In the usual unit $\hbar=c=1$,
with $m$ and ${-e<0}$ mass and charge of the electron one defines usually
$B_{cr}=m^2/e\simeq 4.4\times 10^9\;T.$
In this situation the Landau-level energies are of the order of the electron mass. It is usually thought that around the objects that produce the GRB the magnetic fields are such that $B\ge B_{cr}$; they are slowly varying: if $\omega$, is a frequency typical of the elementary particle dynamics, then
 $|\dot{\mathbf{B}}(t)|/|\mathbf{B}(t)|\ll \omega$.
The production takes place in regions of the order of the Compton wave lengths, so the field may be safely taken as uniform in space: $\mathbf{B}(\mathbf{r},t)\equiv\mathbf{B}(t)$. \par
In view of these features the most suitable scheme of calculation is given by the {\it adiabatic approximation} whose relevant aspect are summarized here below.\par
Given a Hamiltonian $H(\xi)$ with discrete eigenstates, where $\xi=\xi(t)$ is a slowly varying parameter, define:
\begin{align*}
H(\xi)|A,\xi\rangle & =\varepsilon_A(\xi)|A,\xi\rangle\,,\\
H(\xi)|B,\xi\rangle & =\varepsilon_B(\xi)|B,\xi\rangle\,.
\end{align*}
The eigenstate of $H(t_1)$ evolved with $H(t)$ from $t_1$ to $t_2$ is not eigenstate of  $H(t_2)$. There are transitions between  eigenstates of $H(\xi)$ and
the first-order transition amplitude is given by:
\begin{equation*}
\gamma_{BA} (t)=\int_0^t dt'\frac{\langle B,\xi|\partial H/\partial \xi|A,\xi\rangle}{\Delta\varepsilon (\xi)}\dot{\xi}\exp\left[i\int_0^{t'} dt''\Delta\varepsilon (\xi)\right]
\end{equation*}
with
\begin{equation*}
\Delta\varepsilon (\xi)=\varepsilon_B(\xi)-\varepsilon_A(\xi)\,.
\end{equation*}
For the problem we are considering
\begin{align*}
\xi(t) & \longrightarrow \mathbf{B}(t)\,,\\
{|A,\xi\rangle} & {\longrightarrow \small vacuum}\,,\\
{|B,\xi\rangle} & \longrightarrow {\small created\; particles}\,.
\end{align*}
 \section {ELECTRON PRODUCTION}
 We start \cite {1,2,3} by considering a relativistic electron in a constant and uniform magnetic field which lies on the $y$-$z$ plane, choosing a symmetric gauge:
$ \mathbf{A}(\mathbf{r})=
-\frac{1}{2}\left[\mathbf{r}\wedge\mathbf{B}\right] .$
$$ \small
{\cal H}\psi=\left[\pmb{\alpha}\cdot 
\left[-i\pmb{\nabla}+e\mathbf{A}(\mathbf{r})\right]+\beta 
m\right]\psi=\varepsilon \psi .
$$
By using a finite-rotation operator about the $x$ axis acting as:
$
{\cal R}_x(\vartheta){\cal H}{\cal R}_x^{\dag}(\vartheta)=
{\cal H}'$
 the field is aligned with the $z$-direction, the problem is solved once the solutions of the Dirac equation ${\cal H}'\psi'=\varepsilon\psi'\quad$are found.\\
 More in detail 
 \begin{align*}
{\cal H}'\psi'_{\pm,j}(\mathbf{r}) &=\pm
w_j\psi'_{\pm,j}(\mathbf{r})\,,\\
\\
{\cal P}_z\psi'_{\pm,j}(\mathbf{r}) &=k
\psi'_{\pm,j}(\mathbf{r})\,,\\
\\
{\cal J}_z\psi'_{\pm,j}(\mathbf{r}) 
&=\left(n_d-n_g+\frac{s}{2}\right)
\psi'_{\pm,j}(\mathbf{r})\,,\\
\\
R_{xy}^2\psi'_{\pm,j}(\mathbf{r}) &=\frac{2n_g+1}{e\mathrm{B}} 
\psi'_{\pm,j}(\mathbf{r})\,,\\
\\
w_j=&\sqrt{m^2+k^2+e\mathrm{B}(2n_d+1+s)}\,,\\
n_d=0,1,\dots &\qquad n_g=0,1,\dots \qquad s=\pm 1\,.
\end{align*}
 The meaning of the quantum numbers is:
  ${\cal P}_z$: linear momentum along $z$\\
    ${\cal J}_z$: angular momentum along $z$\\
	$R_{xy}^2$: squared distance from $z$ \\
	 Going from electrons to positrons we have the correspondence
$ n_d \leftrightarrow n_g, \quad +s \leftrightarrow -s$
and it is relevant to note that if  $n_d=0, s=-1$ or $n_g=0, s=+1$ then  the energies do not depend on $B$ (transverse ground states).
 \\
 Using these solutions
the relevant operators in second quantization are found
\begin{align*}
H &=\int d\mathbf{r}\;\Psi^{\dag} (\mathbf{r},t){\cal H}\Psi 
(\mathbf{r},t)=\sum_j(w_jN_j+\tilde{w}_j\tilde{N}_j)+E_0\,,\\
P_{\parallel} &=\int d\mathbf{r}\;\Psi^{\dag} 
(\mathbf{r},t){\cal P}_{\parallel}\Psi
(\mathbf{r},t)=\sum_jk(N_j+\tilde{N}_j)\,,\\
J_{\parallel} &=\int d\mathbf{r}\;\Psi^{\dag} 
(\mathbf{r},t){\cal J}_{\parallel}\Psi
(\mathbf{r},t)=\sum_j\left(n_d-n_g+\frac{s}{2}\right)
(N_j+\tilde{N}_j)\,,
\end{align*}
the tilde refers to the antiparticles; $j$ is a multiple index $j\equiv\{n_d,n_g,s,k\}$. \\
The first order amplitude for finding at a time $t$ an $e^--e^+$ pair in the 
state \mbox{$|n_j(t)=1;\tilde{n}_{j'}(t)=1\rangle\equiv |jj'(t)\rangle$} 
from an initial vacuum state   $|0 \rangle$ is: 
 \begin{align*}
\gamma_{jj'}(t)=&\qquad\qquad\\
\int_0^td\tau \frac{\langle 
jj'(\tau)|\dot{H}(\tau)|0(\tau)\rangle}{w_j(\tau)+ 
\tilde{w}_{j'}(\tau)}&\exp\left\{i\int_0^{\tau}d\tau'
\left[w_j(\tau')+ \tilde{w}_{j'}(\tau')\right]\right\} \end{align*}
and the time derivative of the Hamiltonian is 
$$\dot{H}(t)=
-\frac{e}{2} \int d\mathbf{r}\; 
\Psi'^{\dag}(\mathbf{r},t)(\mathbf{r}\wedge 
\pmb{\alpha})\Psi'(\mathbf{r},t)\cdot\dot{\mathbf{B}}'(t)\,.$$
 As anticipated it appears that the energy of the levels grows with $B$, but
  there is a transverse ground level with  energy independent of $B$.
  The angular momentum conservation implies that
  when $\mathbf B$ changes direction both $e^-$ and  $e^+$ can
  be produced in the ground state.
  When $\mathbf B$ changes only in strenght this is not possible so
  the change of direction is more effective than the change of strength.
  For the total rate it results: $ {\cal W} \propto R_{xy}^4. $
  The electric field $\mathbf E$  grows with $R_{xy}$ ({\it Faraday-Neumann} law), so 
$ \Delta {\cal W} /\Delta A \propto R_{xy}^2 \propto E^2$.
 The detailed calculation shows also that there is no unlimited production, but a limiting value, with oscillations in time.\\
 \section{PHOTON PRODUCTION}
 We can foresee three mechanisms of photon production:\\
a- Annihilation of pairs \cite{4}\\
b- Direct creation through nonlinear QED \cite{4}\\
c- Bremsstrahlung\cite{5}.\\
 {\it Annihilation of pairs}\\
 The number of produced photons is the number of produced electron (positrons) times the annihilation cross section times a flux factor:
\begin{align*}
\frac {d{\cal N}(\omega,R_{\perp m},t)}{d\omega dV dt}=&\\
\int\!dkdk' \frac{d\sigma (k,k',\omega)}{d\omega} {\bar v}&
f(k,R_{\perp m},t) f(k',R_{\perp m},t)\,.
\end{align*}
Here $f$ are distributions of produced particles, $d\sigma/d\omega$ the perturbative cross section for annihilation, ${\bar v}$ the flux factor,$R_{\perp m}$ is the mean distance of the produced pair from the origin of the transverse plane. \\
The result is 
a sharp peak in the invariant mass of the photon pair at $2m_e$, this happens
because the main rate of electron production is for longitudinal (along $\mathbf B$) small momenta and (transverse) ground state. In fact
to produce a very soft or very hard photon a large boost is neeeded.\\
{\it Direct creation through nonlinear QED  }\par
 The general form of the effective Lagrangian in nonlinear QED is:\begin{equation*}
{\cal{L}} {={\cal L}({\cal F}_T,{\cal G}_T^2)} \quad
{{\cal F}_T} {=\frac{1}{2}(\mathrm{B}_T^2-\mathrm{E}_T^2)} \quad
{{\cal G}_T} {=\mathbf{E}_T\cdot\mathbf{B}_T}
\end{equation*}
with $\mathbf{E}_T(\mathbf{r},t) =\pmb{\cal{E}}(\mathbf{r},t)$ and $
\mathbf{B}_T(\mathbf{r},t) =\pmb{\cal{B}}(\mathbf{r},t)+\mathbf{B}$.
Here $\pmb{\cal{E}}$ and $\pmb{\cal{B}}$ are the quantized fields and $\mathbf{B}$ is the external magnetic field which rotates with angular velocity $\Omega$.\par
We can choose as effective Lagrangian the well-known Euler-Heisenberg Lagrangian, in this case
for $\rho_0=\mathrm{B}/B_{cr}\gg 1$, but $\alpha\rho_0\ll 1 $ with $\alpha=e^2/4\pi\simeq 1/137$ the fine-structure constant we obtain
\begin{equation*}
\left\langle\frac{d{\cal N}(\omega,t)}{dVd\omega}\right\rangle \sim\frac{28}{135}\frac{\Omega^2}{(2\pi)^3}\left(\frac{\alpha\mathrm{B}}{B_{cr}}\right)^2\,.
\end{equation*}\\
 These results are relevant for low energy photons, anyhow the $e^--e^+$ annihilation is a more efficient mechanism.\\
{\it Bremsstrahlung}\par
 Here the previous formalism must be modified because
 we want to have both the adiabatic production of pairs and the perturbative radiation of real photons.\\
 For rotating $\mathbf B$ we go to a rotating frame by performing a time-dependent rotation
$$ (x,y,z)\to (x',y',z,)=(x, y\cos\Omega t-z\sin \Omega t, z\cos\Omega t +y\sin\Omega t)$$ and rotate accordingly the fields and their time derivatives\\
$e.g.$ \small
$$\frac {\partial\psi(\mathbf{r})}{\partial t} \to \exp\Big[i\frac{\sigma_x}{2}\Omega t\Big] \Big[i\Omega {\cal J}_x \psi'(\mathbf{r}')+\frac {\partial\psi'(\mathbf{r'})}{\partial t}\Big]\,, $$ 
 $${\cal J}_x=-i\bigg(y'\frac{\partial}{\partial z'}-z'\frac{\partial}{\partial y'}\bigg)+\frac{\sigma_x}{2} \quad
{\cal I}_x=-i\bigg(y'\frac{\partial}{\partial z'}-z'\frac{\partial}{\partial y'}\bigg)+S_x\,, $$   \normalsize 
where ${\cal J}_x$ acts on spinor fields, ${\cal I}_x$ on vector fields.\\  This transformation  is not a symmetry transformation!
 It gives a new Lagrangian density (in  powers of $\Omega$)
 \small \begin{align*}
{\cal L}_{\Omega}=
\bar \psi\Big(\gamma^{\mu}\big[i\partial_{\mu}+eA_{\mu}+e{\cal A}_{\mu}\big]-m\Big)\psi-\frac {1}{4}{\cal F}_{\mu,\nu} {\cal F}^{\mu,\nu}\\
+i\Omega \Big[\bar \psi\gamma^o{\cal J}_x\psi-\frac {\partial{\cal A}_{\mu}}{\partial t}{\cal I}_x{\cal A}^{\mu}\Big]+{\cal O}(\Omega^2)\,.
\end{align*}  \normalsize
 Now the external magnetic field is time independent, its time dependence has been shifted to the terms in $\Omega$.\\
A perturbative treatment is built up with
propagators of particles in static $\mathbf {B}$ and two types of
vertices describing either the emission of photons or
the interaction with the $\Omega$ terms.
The treament is analytically complicated, at the end the numerical result is:
\begin{align*}
\frac {dN}{dV\,dt} \propto \frac {1} {\omega} {\rm\; for\; small\;} \omega\,, \\
\frac {dN}{dV\,dt} \propto \frac {1}{\omega^{\beta}} {\rm\;for\;large\;} \omega\,. 
\end{align*}   
The regime changes smoothly around $\omega_b$ depending on $B$, one finds
$\omega_b\approx 1\div 3 $MeV for $B\approx 2\times 10^{10}$ T.  
The exponent found here is $\beta >3$, while the observations prefer $\beta= 2\div 3$.
\section{Effects of gravity}
In order to have huge $\mathbf B$ very compact objects are needed, then relevant effects of gravity are expected.\\
We want to study the case where the gravity acts as a perturbation \cite{6} and also a particular case of strong $i.e.$ nonperturbative gravity\cite{7}. In both cases the gravitational field will be static. We start from the isotropic metric (equivalent to the Schwarzschild metric)
 $$
g_{\mu\nu}(X,Y,Z)=\text{diag} \big[{F^2_-}/{F^2_+},
-F^4_+,-F^4_+,-F^4_+\big]$$
where $F_{\pm}=\pm{r_G}/(4\sqrt{X^2+Y^2+Z^2})$,
 $r_G=2GM$ is gravitational radius of the body and
 $G$ is gravitational constant).\\
{\it The perturbative approach} \\It starts by choosing a point $P_o=(X,Y,Z)$ not too near the horizon and by expanding the metric around it:
\begin{align*}
g^{(1)}_{\mu\nu}(x)=&g^{(0)}_{\mu\nu}+h_{\mu\nu}(x)\,,\\
 g^{(0)}_{\mu\nu} =&
\text{diag}(\phi_t,-\phi_s,-\phi_s,-\phi_s)\,,\\
 h_{\mu\nu}(x) =&
\text{diag}(2g_t x,2g_s x,2g_s x,2g_s x)\,.
 \end{align*}
The coefficients  $\phi$ and $g$ are explicitly given, in terms of $r_G$ and $(X,Y,Z)$; the matrix $h_{\mu\nu}(x)$ is considered as a small perturbation of the zero-order metric tensor  $g^{(0)}_{\mu\nu}$.\\ 
  The perturbed Hamiltonian for the Dirac operators in the presence of a magnetic field is calculated and the new transition amplitudes derived.\\
 There are corrections depending on the potentials $\phi$ 
   and corrections depending on the gradients $g$. 
    Seen from infinity the effect of  $\phi$ is to lower the final energy.\\
{\it A nonperturbative example} \\
Now the pair is created  near the black hole event horizon lying at 
$\sqrt{X^2+Y^2+Z^2}=r_G/4$.
We start by expanding the metric tensor 
$g_{\mu\nu}$ around the point $P_o=(0,0,r_G/4)$. 
In so doing the metric is reduced to a Rindler metric for which the solution of the Dirac equation, in a magnetic field parallel to the gravitational field is known.
Two results of this calculation:\\
1- there are pair that cannot fly at infinity, they give (also) low-energy photons {\it (expected)}.\\
2- the dependence on the strength of $B$ is stronger than in Minkowski space
{\it (not obvious)}.\\
\section {EXOTICS}
Till now the production of electrons and photons has been investigated,
but in principle the same mechanism can give rise to other particles, we simply
list and brifly comment the various possibilities:\\
Production of $\mu^{\pm}$, the calculation are the same as for the  $e^{\pm}$,
the rate is very small.\\
Production of $\pi^{\pm}$, the calculation is not very different,
the rate is even smaller, since there are no states with energy independent of
$\bf B$.\\
Production of $\pi^{o}$ \cite{8}:  it would be possible due to a mixing, in very strong
magnetic field, of $\pi^{o}$ and $\rho^{o}$, the rate is even smaller than for
charged pions.\\
Production of neutrinos \cite{4}: if the neutrinos have mass they can have anomalous
magnetic moment, but the size of it is controlled by the vector-meson masses, so
the production is very small, in spite of the small mass.\\
Production of axions \cite{9}: assume the phenomenological linear coupling $C\Phi\;\bf {E\cdot B}$. This form of coupling yields a production which depends also on the spatial variation of the fields and a number distribution which is Poissonian for every definite mode.\\
 
\noindent{\bf Acknowledgements:} 
Presented at ISMD2006, Paraty (R.J.) Brazil, September 2006.
\par
 In order to save space only the papers used in producing this presentation
 have been quoted, in them the relevant references can be found.

\vfill
\end{document}